# Extreme mobility enhancement of two-dimensional electron gases at oxide interfaces via charge transfer induced modulation doping

Y. Z. Chen[1*], F. Trier,[1] T. Wijnands,[2] R. J. Green,[3,10] N. Gauquelin,[4] R. Egoavil,[4] D. V. Christensen,[1] G. Koster,[2] M. Huijben,[2] N. Bovet,[5] S. Macke,[3,11] F. He,[6] R. Sutarto,[6] N. H. Andersen,[7] J. A. Sulpizio,[8] M. Honig,[8] G. E. D. K. Prawiroatmodjo,[9] T. S. Jespersen,[9] S. Linderoth,[1] S. Ilani,[8] J. Verbeeck,[4] G. Van Tendeloo,[4] G. Rijnders,[2] G. A. Sawatzky,[3] N. Pryds[1]

*[1]Department of Energy Conversion and Storage, Technical University of Denmark, Risø Campus, 4000 Roskilde, Denmark*

*[2]Faculty of Science and Technology and MESA+ Institute for Nanotechnology, University of Twente, 7500 AE Enschede, The Netherlands.*

*[3]Quantum Matter Institute, Department of Physics and Astronomy, University of British Columbia, Vancouver, British Columbia V6T 1Z1, Canada.*

*[4]EMAT, University of Antwerp, Groenenborgerlaan 171, 2020 Antwerp, Belgium.*

*[5]Nano-Science Center, Department of Chemistry, University of Copenhagen, 2100 Copenhagen, Denmark*

*[6]Canadian Light Source, University of Saskatchewan, Saskatoon, SK S7N2V3, Canada.*

*[7]Department of Physics, Technical University of Denmark, 2800 Lyngby, Denmark*

*[8]Department of Condensed Matter Physics, Weizmann Institute of Science, 76100 Rehovot, Israel.*

---

[*] Correspondence and requests for materials should be addressed to Y.Z.C (Email: yunc@dtu.dk).

[9]*Center for Quantum devices, Niels Bohr Institute, University of Copenhagen, 2100 Copenhagen, Denmark*

[10]*Max Planck Institute for Chemical Physics of Solids, Nöthnitzerstraße 40, 01187 Dresden, Germany.*

[11]*Max Planck Institute for Solid State Research, Heisenbergstraße 1, 70569 Stuttgart, Germany.*

**The discovery of two-dimensional electron gases (2DEGs) at the interface between two insulating complex oxides, such as LaAlO$_3$ (LAO) or gamma-Al$_2$O$_3$ (GAO) epitaxially grown on SrTiO$_3$ (STO) [1,2], provides an opportunity for developing all-oxide electronic devices[3,4]. These 2DEGs at complex oxide interfaces involve many-body interactions and give rise to a rich set of phenomena[5], for example, superconductivity[6], magnetism[7,8], tunable metal-insulator transitions[9], and phase separation[10]. However, large enhancement of the interfacial electron mobility remains a major and long-standing challenge for fundamental as well as applied research of complex oxides[11-15]. Here, we inserted a single unit cell insulating layer of polar La$_{1-x}$Sr$_x$MnO$_3$ ($x$=0, 1/8, and 1/3) at the interface between disordered LaAlO$_3$ and crystalline SrTiO$_3$ created at room temperature. We find that the electron mobility of the interfacial 2DEG is enhanced by more than two orders of magnitude. Our *in-situ* and resonant x-ray spectroscopic in addition to transmission electron microscopy results indicate that the manganite layer undergoes unambiguous electronic reconstruction and leads to modulation doping of such atomically engineered complex oxide heterointerfaces. At low temperatures, the modulation-doped 2DEG exhibits clear Shubnikov-de Haas oscillations and the initial manifestation of the quantum Hall effect, demonstrating**

**an unprecedented high-mobility and low electron density oxide 2DEG system.
These findings open new avenues for oxide electronics.**

One of the most important developments in semiconductors, both from the viewpoint of fundamental physics and for the purpose of developing new devices, has been the realization of 2DEGs in heterointerfaces based on Si or on III–V compounds. The quality of the material hosting the 2DEG is quantified predominantly by its carrier mobility, which is limited due to the scattering of electrons by ionized impurities, particularly at low temperatures[16,17]. The usage of the so-called modulation-doping technique, artificially separating the electrons from the ionized impurities by inserting a spacer at the interface, has significantly increased the carrier mobility of 2DEGs in semiconductor heterostructures[16-18]. This led not only to the realization of high-electron-mobility field effect transistors but also to the discovery of the fractional quantum Hall effect in 2DEG samples of unprecedented structural perfection[19].

The enhancement of 2DEG carrier mobilities at complex oxide interfaces, by contrast, remains rather challenging. To date, one of the most investigated complex oxide 2DEGs is based on interfacing epitaxial LAO and STO (*c*-LAO/STO), where both electronic reconstruction[1, 20, 21] and ion transfer across the interface[22, 23] can make important contributions to the conduction. In a different but related system, the interface between disordered LAO and STO (*d*-LAO/STO) created at room temperature holds a 2DEG resulting from interfacial redox reactions[24-26]. For both types of STO 2DEGs, they exhibit a distinct electronic feature: the interfacial charge carriers ($t_{2g}$ electrons) come in two species, $d_{xy}$ and $d_{xz}/d_{yz}$ (z is perpendicular to the 2D layers)[27-30]. The $d_{xy}$ component accounts for most of the 2DEG charge. They reside at or in the immediate proximity of the interface, but they exhibit a rather low mobility. In contrast, $d_{xz}$ or $d_{yz}$ electrons naturally extend further away from the interface and amount to only a small fraction of the 2DEG population. Yet, their mobilities reach much higher values than that of the $d_{xy}$ states[27-30].

So far, there are experimental indications that decreasing the charge carrier density could enhance the electron mobility[12-15]. However, the experiments to control the charge carriers have mainly focused on either controlling processes at the surface of the heterostructures or by simply reducing the contributions of defect dopants such as oxygen vacancies[12-15]. Considering the electronic structure as well as the inherent lack of long-range translational symmetry in the $d$-LAO/STO system, which could give rise to random potentials across the interface leading to low mobilities[24], in this study, we performed investigations to tackle both the issues of suppressing the charge carrier density as well as improving the interface order by introducing an additional layer at the interface of $d$-LAO/STO. The buffer layer is designed to serve three purposes: 1) to reduce the 2DEG carrier density by selecting a material which can trap electrons in a controlled manner and therefore enhance the carrier mobility, 2) to improve the interfacial order and 3) to act as a separating layer between the conducting layer and its source(s) of doping. Most perovskite oxides that have a prominent characteristic of lower Fermi level compared to STO 2DEGs can be good candidates.

Here, we report a 2DEG mobility enhancement of more than two orders of magnitude in $d$-LAO/STO obtained by inserting a single unit cell (uc) buffer layer of manganite[31], $La_{1-x}Sr_xMnO_3$ (LSMO, $x$=0, 1/8, and 1/3), at the interface, as illustrated in Fig. 1a. As it will become clear, this leads to an unexpected modulation-doping scheme of the complex oxide 2DEG, and results in a very high 2DEG mobility exceeding 70 000 $cm^2V^{-1}s^{-1}$ at 2 K. The fact that the final step to create these complex oxide 2DEGs occurs at room temperature, offers a notable advantage of improved compatibility with the established lithography of semiconductor microfabrication to pattern oxide interfaces[15, 32]. This opens new doors for oxide nanoelectronic devices such as tunnel junctions and high mobility field-effect transistors[33].

Our samples were grown at room temperature on $TiO_2$-terminated single-crystal STO substrates by pulsed laser deposition. For the buffered sample, an epitaxial LSMO layer was first deposited at 600 ºC and $1\times10^{-4}$ mbar of $O_2$, then slowly cooled to room temperature at the approximate deposition pressure ($10^{-5}$-$10^{-4}$ mbar) before switching to *d*-LAO deposition *in situ*. The LSMO layer with a thickness of 1 uc was accurately controlled by monitoring *in situ* reflection high-energy electron diffraction (RHEED) intensity oscillations under optimized condition of layer-by-layer epitaxial film growth mode (Supplementary Information, Fig. S1). For all samples, with the *d*-LAO film thickness up to 20 nm, atomic force microscopy (AFM) images show regular surface terraces with a step height of 0.4 nm (Fig. 1b), similar to that of the original STO substrate and indicating uniform film growth. Representative samples were investigated further by scanning transmission electron microscopy (STEM). Figure 1c shows high-angle annular dark field (HAADF) STEM images of a LSMO ($x$=1/8) buffered sample with the *d*-LAO layer of approximately 8.0 nm. The room temperature deposited and un-annealed LAO film is found to be disordered or amorphous-like, and the nominally 1 uc LSMO film is found to be coherently grown on the STO substrate with no detectable defects or dislocations at the LSMO/STO interface. We observe what is likely an intrinsic crystallization of the LaO sublayer in close proximity to the LSMO buffer layer during room temperature film growth of LAO (Supplementary Information, Figs. S2-S4). The first layer of the subsequent *d*-LAO might be partially crystallized in the form of islands according to the contrast differences within this layer, but accurate quantification of the amorphous/crystalline ratio from such a projected image remains open[34,35]. The further investigation of chemical composition across the interface by electron energy-loss spectroscopy (EELS) confirms the interfacial confinement of Mn, in addition to limited cation intermixing within 0.8 nm (Supplementary Information, Figs. S2 and S4). Note that EELS spectra at the Mn-$L_{2,3}$ edge indicate a much reduced manganese valence compared to the norminal $Mn^{3+}/Mn^{4+}$ of bulk LSMO (Fig. 1d). Additionally, the possibility of La-doping of STO induced interface conduction is ruled out here, because all our LSMO/STO heterostructures with LSMO up to 4 uc are highly insulating without the growth of the capping $d$-LAO layer.

The presence of the LSMO buffer layer is found to have a strong impact on the transport properties of the $d$-LAO/STO structure. Firstly, when the buffer layer thickness, $t$, is increased from 1 uc to 2 uc, a metal-to-insulator transition is observed (Figs. 2a and b), i.e. the buffered heterostructures turn highly insulating when $t \geq 2uc$ (see also Supplementary Information, Fig. S5). Secondly, in the case with a single unit cell buffer layer, the transport behaviour of these room-temperature created STO-based heterostructures is improved greatly. Figures 2c-e show the typical electrical transport properties of $d$-LAO/STO interfaces with and without the single unit cell LSMO buffer layer. The unbuffered $d$-LAO/STO interface shows a good metallic behavior with electron density, $n_s$, of approximately $1.2 \times 10^{14}$ cm$^{-2}$ at 300 K. However, upon cooling, it always exhibits a carrier freeze-out effect at $T \leq 100$ K with an activation energy of 5-10 meV[26], indicating the presence of localized electrons. This results in a low electron mobility, $\mu$, of approximately 600 cm$^2$V$^{-1}$s$^{-1}$ at 2 K. In contrast, all LSMO buffered samples exhibit a nearly temperature independent $n_s$ in addition to a striking decrease in sheet resistance, $R_s$, of about 3-4 orders of magnitude during cooling. This suggests the removal of localized electrons and improved metallic conduction by the insertion of LSMO layers. Furthermore, all metallic LSMO buffered samples exhibit a strongly suppressed $n_s$ in the range of $3.4\times10^{12}$ cm$^{-2}$ to $2.7\times10^{13}$ cm$^{-2}$ (0.005-0.040 electrons per uc), nearly independent of $x$. With respect to electron mobility, more than 30 samples have been grown with high mobilities exceeding 10000 cm$^2$V$^{-1}$s$^{-1}$ at 2 K, which is more than 10 times higher than mobilities of unbuffered samples. Representatively, at 2 K, a $n_s=6.9\times10^{12}$ cm$^{-2}$, $\mu=16000$ cm$^2$V$^{-1}$s$^{-1}$, a $n_s=1.7\times10^{13}$ cm$^{-2}$, $\mu=73000$ cm$^2$V$^{-1}$s$^{-1}$ and a $n_s=1.8\times10^{13}$ cm$^{-2}$, $\mu=8800$ cm$^2$V$^{-1}$s$^{-1}$, are obtained for $x=0$, 1/8 and 1/3, respectively. The very high electron mobility of 73000 cm$^2$V$^{-1}$s$^{-1}$ is larger than any-yet reported mobilities for 2DEGs based on the LAO/STO system[3-14].

Our subsequent spectroscopic measurements reveal dramatic electronic reconstruction in the LSMO-buffered samples. Firstly, different from the unbuffered *d*-LAO/STO sample where the 2DEG is coupled strongly to a large content of oxygen vacancies extending more than 3 nm deep into STO[24], all buffered samples show a rather low content of Ti$^{3+}$ far below the detection limit of our *in situ* X-ray photoelectron spectroscopy (XPS) measurements (Supplementary Information, Fig. S6). This suggests much reduced carrier density and/or oxygen vacancies compared to the unbuffered sample, consistent with the transport measurements (Figs. 2c-e). Similar results are also detected by our spatially resolved EELS measurements which discover a nearly perfect Ti$^{4+}$ for the LMO buffered case and some remaining Ti$^{3+}$ confined exclusively to the first two unit cells (~0.8 nm) of STO in the LSMO ($x=1/8$) buffered case (Supplementary Information, Fig. S7). Secondly, both *in-situ* XPS (Fig. S6) and EELS spectra (Fig.1D and Fig.S7) data show an unambiguous signature of Mn$^{2+}$ in the buffer layer, which appears only when the *d*-LAO film is thicker than 2 nm (Fig. S6). The lowering of the Mn valence and its impact on the 2DEG carriers in the buffered samples are further determined by synchrotron radiation based resonant X-ray reflectivity (RXR) (Supplementary Information, Fig. S8), which can probe buried interfaces and thin layers with extremely high sensitivity.

Figures 3a and b show the reflectivity measurements across the Ti and Mn $L_{2,3}$ resonances, respectively, for both unbuffered and the LSMO ($x$=1/8) buffered samples. The overall shapes of the spectra across the Ti $L_{2,3}$ resonance (Fig. 3a) are indicative of the $Ti^{4+}$ in the STO with distinct stronger $e_g$ peaks relative to the $t_{2g}$ peaks at both the $L_3$ (~460 eV) and $L_2$ (~466 eV) edges for the unbuffered sample. This confirms the stronger presence of $Ti^{3+}$ (i.e. occupied 2DEG $t_{2g}$ states) in the unbuffered sample as independently determined by both XPS and EELS. The enhancement near the $e_g$ peak as a response to the 2DEG is also found in resonant photoemission spectra from $c$-LAO/STO samples[36]. In strong agreement with the EELS and XPS spectra, the RXR scan across the Mn resonance (Fig. 3b) also reveals the presence of $Mn^{2+}$ in the buffer layer. Remarkably, an X-ray absorption spectra (XAS) calculation based on a theoretical cluster model (Supplementary Information, Table S1), which can be comparable to the experimental RXR spectra, finds that the experimental spectrum is due to the presence of almost entire $Mn^{2+}$, exhibiting a relatively weak $e_g$ - $t_{2g}$ crystal field splitting energy about one half in magnitude of that in $MnO^{37}$. This presence of nearly entire $Mn^{2+}$ indicates an extremely high concentration of reconstructed electrons of ~1 $e$/uc in the LSMO buffer layer, much higher than the deduced $n_s$ of the 2DEG (Fig. 2). This implies that the reconstructed electrons at the $Mn^{2+}$ site are mostly trapped as expected, while the 2DEG carriers stay on the Ti site as in the case of the unbuffered sample. Moreover, the buffered samples show a total amount of reconstructed electrons (~1.005 -1.040 e per uc square area) much higher than that of the unbuffered sample (typically below 0.5 e per uc square area[24]). This dramatic difference probably comes from the polarity of the LSMO buffer layer and the $d$-LAO film. Note that, though without any detectable long-range translational symmetry, the $d$-LAO locally could contain $AlO_6$ octahedra that are similar to the local bonding units in $c$-LAO. Moreover, the polarity of $d$-LAO can be tuned by an external dipole[38]. Therefore, it is reasonable that the polar nature of the LSMO buffer layer is inherited or even enhanced in the $d$-LAO system, and that the polarization catastrophe induced electronic reconstruction[20] can well explain the electron sources in the buffered sample. We further confirmed that the reduction of LSMO buffer layer during the $d$-LAO deposition occurs hardly as indicated by another control experiment: an insulating behaviour and negligible signature of $Mn^{2+}$ was detected when the bilayer of $d$-LAO/LSMO was grown on a $(LaAlO_3)_{0.3} (Sr_2AlTaO_6)_{0.7}$ substrate under the exactly same condition as the buffered $d$-LAO/STO sample, consistent with the previous report[39]. Note that we can not rule out the possibility that some oxygen vacancies might be present in the buffer layer due to its oxygen-deficient growth environment (mainly for $x$=1/8 and 1/3), though their amount is expected to be quite small as determined by XPS before the $d$-LAO deposition (Supplementary Information, Fig. S6). Due to this possible presence of oxygen vacancies in the buffer layer, it becomes challenging to compare the buffered samples with different Sr content ($x$), which also suffer from various strain effects. Nevertheless, by comparing all the buffered samples with the unbuffered one, we can conclude unambiguously that, the manganite buffer layer prevents the reduction of STO and induces a remarkable electronic reconstruction: most of the interfacial electrons, which otherwise stay on Ti are transferred to Mn and become highly trapped.

The effective electron absorption by LSMO in the buffered $d$-LAO/STO results from the unique band alignment of the manganite-titanate perovskite interface. The undoped LaMnO$_3$ (LMO) is a charge transfer insulator. Its electronic configuration of the Mn $d$ electrons is $t_{2g}^3 e_g^1$. Generally, a static Jahn-Teller distortion lifts the degeneracy in the $e_g$ level, splitting it into $e_{g1}$ and $e_{g2}$ levels with the energy separation between $e_{g1}$ and the empty $e_{g2}$ levels being about 1.9 eV[40]. The substitution of La$^{3+}$ by Sr$^{2+}$ ions produces holes in the $e_g$ orbitals which are strongly hybridized with oxygen 2$p$ states. As the doping level $x$ is increased, a downward shift of the Fermi level position is expected. On the other hand, the Fermi level of the STO 2DEG is close to that of Nb-doped STO. When manganites are grown epitaxially on the titanate, the interfacial barrier height ($\Phi_B$) between the slightly electron-doped STO and the LSMO is determined to be increased from approximately 0.5 eV to 1.0 eV when $x$ increases from 0 to 0.5[41,42]. Therefore, the Ti 3$d$ bands of STO are always higher than the empty or partially filled $e_g$ bands of LSMO, given the alignment of O 2$p$ bands when they come into contact, as illustrated in Fig. 4a. In this vein, during the formation of 2DEG in buffered samples, electrons, which originate from positively ionized donors on the $d$-LAO side, will be firstly transferred to the Mn sublattice before filling the electronic shell of Ti ions. This scenario is consistent with the fact that Mn$^{2+}$ becomes detectable already at an $d$-LAO film thickness of 2.0 nm (Fig.S6), much before the critical thickness for occurrence of interface conduction, ~3.2 nm for the $x$=1/8 buffered sample (Fig. S5). Moreover, the above electron transfer process is expected to modulate significantly the band alignment at buffered $d$-LAO/STO interfaces. Taking the LMO buffer layer as an example, transferred electrons will fill completely the first empty subband ($e_{g2}^\uparrow$) of LMO (Fig. 4a). This results in a nontrivial consequence that the bottom of the conduction band ($t_{2g}^{\downarrow}$) of the reconstructed buffered layer becomes higher than that of the capping $d$-LAO (Fig. 4b). Consequently, the LSMO buffer layer, remaining highly insulating both before and after electron transfer, could act as a spacer as introduced in the modulation doped semiconductor heterostructures[16,18] (Supplementary Information, Fig. S9) that spatially separates the 2DEG electrons (on the STO side) and the ionized donors (on the $d$-LAO side). Therefore, the buffered samples exhibit a much higher mobility. Interestingly, such band alignment of the buffered sample, particularly the filling of the $e_g$ subbands by charge transfer, is strongly supported by our spatially resolved EELS spectra at the O-K edges (Supplementary Information, Figs. S10 and S11). Moreover, the electronic reconstruction scheme is also consistent with the insulating nature of the buffered sample when the LSMO buffer layer is increased to 2 uc (Figs. 2a and b): The total reconstructed electrons resulting from interface polarity are expected to be constant regardless of the LSMO thickness. When the buffer layer is thicker than 1 uc, the reconstructed electrons can not fill fully the empty subbands of LSMO, therefore there will be no electrons transferred to the STO conduction bands thus no formation of 2DEGs.

Finally, it is noteworthy that the combination of high mobility and low carrier density of our modulation-doped 2DEGs may enable the observation of the quantum Hall effect. Here, the electrons in our modulation-doped 2DEGs with a true 2D Fermi surface, i.e. constant density of states (DOS) as a function of energy, and a sufficiently high mobility form extended edge states at high magnetic fields. At these fields the DOS becomes discretized into separate Landau Levels with an integer filling factor. For sufficiently low carrier density 2DEGs these Landau Levels become experimentally resolvable at available fields and low temperatures. Fig. 5a shows transport measurements at $T$=30 mK of the longitudinal resistance, $R_{xx}$, and the transverse resistance, $R_{xy}$, as a function of magnetic field ($B$) applied perpendicular to a Hall-bar sample with $n_s$=5.5×10$^{12}$ cm$^{-2}$. In spite of a compromise in the electron mobility of our patterned oxide interface, probably due to processing, $R_{xx}$ shows clear Shubnikov-de Haas (SdH) oscillations.[44] By subtracting a magnetoresistance background, the oscillations are periodic in 1/$B$ and dominated by a single oscillation frequency as shown in Fig. 5b. This behavior is similar to other STO-based conduction heterostructures with extremely high mobility.[2,45] At fields where the $R_{xx}$ shows a minimum (indicated by vertical dashed lines), there is a corresponding reduction of the $R_{xy}$ slope consistent with the initial formation of quantized Hall plateaus.[43] The observation of quantum Hall-like structure in our 2DEG further supports the feasibility of our modulation doping approach and establishes the quality of our samples. Detailed analysis of the density, temperature and angle dependence of the quantum Hall results will be presented elsewhere.[46]

In conclusion, the LSMO buffer layer not only suppresses significantly the oxygen vacancies on the STO side but also results in modulation doping in buffered $d$-LAO/STO, which remains underexploited at complex oxide interfaces.[11] Since most of complex oxide 2DEGs hold a similar electronic structure[27-30], the charge transfer induced modulation doping discussed here is expected to be a universal phenomenon. It could result in a plethora of exotic physical phenomena and therefore represents a milestone for oxide electronics.

**Methods**

**Sample growth and transport measurements**

All heterostructures were grown on TiO$_2$-terminated SrTiO$_3$ (STO) substrates (5×5×0.5 mm$^3$ with miscut less than 0.2º) by pulsed laser deposition (PLD) in an oxygen atmosphere of ~10$^{-4}$ mbar with the film growth process monitored by *in-situ* RHEED. During ablation, a KrF laser (λ=248 nm) with a repetition rate of 1 Hz and laser fluence of 1.5 Jcm$^{-2}$ was used. The target-substrate distance was fixed at 5.6 cm. For disordered LaAlO$_3$ (*d*-LAO) films deposited at room temperature at 10$^{-5}$-10$^{-4}$ mbar, a single crystalline LAO target was used. For the buffered La$_{1-x}$Sr$_x$MnO$_3$ (LSMO, *x*=0, 1/8, 1/3) layers, sintered LSMO ceramics were used as targets. The growth temperature of LSMO was fixed at 600 ºC. Layer-by-layer two-dimensional growth of LSMO films was optimized by RHEED oscillations with a growth rate of approximately 50 pulses per unit cell (~4.8 Å/min, Supplementary Information, Fig. S1a). For each buffered sample, after the epitaxial growth of monolayer LSMO, the sample was cooled under the deposition pressure with a rate of 15 ºC/min to room temperature (below 30 ºC, in 5-6 hours) before the subsequent *d*-LAO film deposition *in situ* (Fig. S1d). Note that *in situ* annealing of our samples at 80 ºC in 10$^{-2}$ mbar oxygen pressure results in insulating samples, indicating that the ionized donors may be strongly coupled to oxygen vacancies at the sample surface.

Electrical characterization was made mainly using a 4-probe Van der Pauw method with ultrasonically wire-bonded aluminum wires as electrodes. Samples patterned in a Hall-bar configuration also showed similar enhanced electron mobilities. All buffered samples exhibit linear Hall resistance in magnetic fields.

Quantum transport measurements were performed on a patterned Hall bar device (*W*=50 *μ*m, *L*=300 *μ*m) in a dilution refrigerator at 30 mK with magnetic fields up to 12 T using a standard lock-in technique.

**STEM and EELS analysis**

Transmission electron microscopy studies were performed at room temperature using the Qu-Ant-EM microscope which is an aberration corrected scanning transmission electron microscope (STEM), FEI Titan3 80-300 operated at an acceleration voltage of 200 kV, equipped with a high-brightness field emission electron source (X-FEG) and a high-resolution electron energy-loss spectrometer. Cross-sectional cuts of the samples grown under the optimum condition described above were prepared using a dual-beam focused ion beam (FIB) instrument. For high angle annular dark field (HAADF) imaging, a probe size of approx. 1 Å with a convergence angle of 24 mrad, and an HAADF inner collection angle of 140 mrad were used. For electron energy-loss spectroscopy (EELS) in STEM, the collection angle was set to 36 mrad. Due to the very high sensitivity of the *d*-LAO layer, the challenge for atomically resolved EELS and STEM imaging of the heterostructures is to minimize radiation damage resulting in recrystallization of the *d*-LAO layer. This was taken care of through the use of a very low electron dose for both STEM imaging and EELS spectroscopy (< 40pA). The acquisition time for EELS could not be set to more than 0.05 sec/pixel with a step size of 0.2 Å/pixel resulting in rather low signal-to-noise ratio and making it very difficult to extract the small signal of Mn from the spectrum image. Two-dimensional spectrum images were acquired with STEM-EELS to investigate the spatial distribution of the elements within the sample. To improve the signal to noise ratio (SNR) in the EELS spectra (mainly on the Mn signal), principal components analysis (PCA) was applied. This analysis technique is a powerful tool to reduce the noise from STEM-EELS data which enables one to extract the fundamental chemical information[48,49]. The elemental maps were generated by subtracting and integrating the corresponding core-loss excitation edge for each chemical element and are presented in figures S2a and S4a. Another way to improve the SNR in the EELS spectra is as follows: Using the raw data, after subtraction of the corresponding power law background, EELS spectra on each line are summed and the result is plotted as a function of the vertical direction of the spectrum image (growth direction). The edges are then plotted as a contour plot in a color scale as shown in Figures S2b and S4b giving direct insight on compositional and electronic structure changes at atomic resolution.

*In-situ* **X-ray photoelectron spectroscopy (XPS) measurement**

For the study using angular resolved XPS, a series of samples were first grown by PLD. Subsequently, the samples were transferred to the XPS chamber while keeping them under ultra high vacuum (below $1\times10^{-9}$ mbar). The XPS chamber (Omicron Nanotechnology GmbH) had a base pressure below $1\times10^{-10}$ mbar. The measurements were done using a monochromatic Al Kα (XM 1000) X-ray source and an EA 125 electron energy analyzer. All spectra were acquired in the Constant Analyzer Energy (CAE) mode. A CN 10 charge neutralizer system was used to overcome the charging effect in the LSMO/STO heterostructures. For each measurement the filament current, emission current and beam energy were optimized to minimize the full width at half maximum (FWHM) of the peaks. For analyzing the Ti $2p_{3/2}$ peaks, a Shirley background was subtracted and the spectra were normalized on the total area below the Ti peaks ([Ti] = [Ti4+] + [Ti3+] = 100%). To investigate the dependence of interface states on the film thickness of *d*-LAO, the XPS measurement was performed after each 2 nm deposition of *d*-LAO.

**Synchrotron radiation resonant x-ray reflectivity (RXR) experiments**

The RXR measurements were performed using an in-vacuum 4-circle diffractometer at the Resonant Elastic and Inelastic X-ray Scattering (REIXS) 10ID-2 beamline of the Canadian Light Source (CLS) in Saskatoon, Canada[50]. The beamline has a flux of $5\times10^{12}$ photon/s and photon energy resolution $\Delta E/E$ of $\sim10^{-4}$. The base pressure of the diffractometer chamber was kept lower than $10^{-9}$ Torr. The samples were aligned with their surface normal in the scattering plane. All data reported here were collected at room temperature (293 K) with σ-polarized light (polarization perpendicular to the scattering plane). The measurements were carried out in the specular reflection geometry with either constant photon energies or constant momentum transfer across Ti $L_{2,3}$ resonances (~450-470 eV) and Mn $L_{2,3}$ resonances (~635-660 eV). The experimental geometry is depicted in Fig. S8.

## Acknowledgements

The authors gratefully acknowledge the discussions with J. Mannhart, J. R. Sun and B. G. Shen, and the technical assistance from J. Geyti, L. Han, K.V. Hansen, S. Upadhyay, C. Olsen, and A. Jellinggaard. This work was funded by the European Union (EU) Council under the 7th Framework Program (FP7) grant nr NMP3-LA-2010-246102 IFOX, by funding from the European Research Council (ERC) under FP7, ERC grant N°246791 - COUNTATOMS and ERC Starting Grant 278510 VORTEX. The Qu-Ant-EM microscope was partly funded by the Hercules fund from the Flemish Government. The authors acknowledge also financial support from EU under FP7 under a contract for an Integrated Infrastructure Initiative. Reference No. 312483-ESTEEM2. Funding from the fund for scientific research flanders is acknowledged for FWO project G.0044.13N ('Charge ordering'). Funding from the Danish Agency for Science, Technology and Innovation, and the Lundbeck Foundation are acknowledged. The Center for Quantum Devices is supported by the Danish National Research Foundation. The Canadian work was supported by NSERC and the Max Planck-UBC Centre for Quantum Materials. Some experiments for this work were performed at the Canadian Light Source, which is funded by the Canada Foundation for Innovation, NSERC, the National Research Council of Canada, the Canadian Institutes of Health Research, the Government of Saskatchewan, Western Economic Diversification Canada, and the University of Saskatchewan.

## Author Contributions

Y.Z.C. concept design, film growth, transport measurements, data analysis, interpretation and writing of the manuscript. The contributions of other authors are as follows. Transport measurements and analysis: D.V.C., N.H.A., M.H., J.A.S., M.H., S.I.; STEM and EELS measurements and analysis: N.G., R.E., J.V., G.V.T.; XPS measurements and analysis: T.W., G. K., M.H., G.R., N.B.; RXR measurements and analysis: R.J.G., S.M., F.H., R.S., G.A.S; SdH measurements: F.T., G.E.D.K.P., T.S.J.; Data discussion: N.P., S.L. All authors extensively discussed the results and the manuscript.

## Additional information

The authors declare no competing financial interests. **Supplementary information** accompanies this paper on www.nature.com/naturematerials. **Reprints and permissions** information is available online at http://npg.nature.com/reprintsandpermissions. Correspondence and requests for materials should be addressed to Y.Z.C.

**Figure legends:**

**Figure 1 Atomically engineered complex oxide interfaces with a single unit cell manganite buffer layer. a**, Sketch of the single unit cell LSMO-buffered $d$-LAO/STO heterostructures. **b**, AFM image (2.5 $\mu$m×5 $\mu$m) of a 20 nm $d$-LAO/1 uc LSMO ($x$=1/8)/STO heterostructure showing an atomically smooth surface. **c**, HAADF-STEM images of a 8 nm $d$-LAO/1 uc LSMO ($x$=1/8)/STO heterostructure. The brighter two LaO layers (marked with white dotted lines) determine the 1 uc LSMO layer. The occasional LaO islands probably result from intrinsic epitaxial crystallization of LAO at the disordered/crystalline interface during room-temperature film growth. Spatially resolved Mn $L_{2,3}$ EELS spectra, **d**, confirm the confinement of the LSMO buffer layer at the interface. The $L_3$ peak position at 640 eV and the $L_2$ features indicated by arrows are characteristic of $Mn^{2+}$ states in good agreement with other spectroscopic results (Fig. 3). More details are presented in the supplementary information (Figs. S2-S4).

**Figure 2 Electronic properties of $d$-LAO (8.5 nm)/STO heterostructures with and without LSMO ($x$=0, 1/8, and 1/3) buffer layers. a** and **b**, Sheet conductance, $\sigma_s$, and the carrier density, $n_s$, respectively, as a function of the buffer layer thickness, $t$, for LSMO ($x$=1/8) buffered $d$-LAO/STO hetero-structures. All data were obtained at room temperature. **c-e**, The temperature dependence of sheet resistance, $R_s$, carrier density, $n_s$, and mobility, $\mu$, respectively. The unbuffered $d$-LAO/STO always shows a carrier freeze-out at around 100 K, while, the buffered samples exhibit a remarkable temperature

independent $n_s$ and a large $\mu$ often exceeding approximately 10000 cm$^2$V$^{-1}$s$^{-1}$ at 2 K. The lines are guides to the eye.

**Figure 3 Electronic reconstructions in *d*-LAO/STO heterostructures. a** and **b,** the RXR fixed-momentum measurements with photon energies scanned across the Ti and Mn $L_{2,3}$ edges, respectively. The insets show momentum scans at fixed energies of 459.5 eV and 640.3 eV for Ti and Mn respectively, as indicated by the vertical dashed lines. Vertical bars in the insets show the specific momentum values used for the RXR measurements. The interference fringes are primarily defined by the *d*-LAO layer thickness (8.5 nm). Mn X-ray absorption spectra (XAS) of reference materials MnO (Mn$^{2+}$), LaMnO$_3$ (Mn$^{3+}$), and SrMnO$_3$ (Mn$^{4+}$)[47] are shown in **b** for comparison, as well as a calculated spectrum of Mn$^{2+}$.

**Figure 4 Modulation doping of STO-based heterostructures. a** and **b**, Schematic band diagram of LMO-buffered *d*-LAO/STO before and after the *d*-LAO deposition, respectively. Given the alignment of O *2p* bands when STO and LMO become contact, Ti *3d* bands are higher than the empty or partially filled $e_g$ bands of LSMO, as illustrated in **a**. Therefore, reconstructed electrons coming from the positively ionized donors will be transferred firstly to LSMO buffered layer. The formation of 2DEG occurs only after the presence of Mn$^{2+}$(when *d*-LAO is thicker than 3.2 nm), as shown in **b**.

**Figure 5 Quantum oscillations at modulation-doped oxide interfaces. a,** Longitudinal resistance, $R_{xx}$, and Hall resistance, $R_{xy}$, as a function of magnetic field *B* normal to the sample surface. $R_{xx}$ exhibits Shubnikov-de Hass (SdH) quantum oscillations with minima at the same positions as plateau-like

structures in $R_{xy}$, indicating the observation of the quantum Hall effect. **b**, Amplitude of the SdH oscillations, $\Delta R_{xx}$, shows $1/B$ periodicity, which is dominated by a single band conduction with the frequency of the oscillations approximately 21 T.

**Figures:**

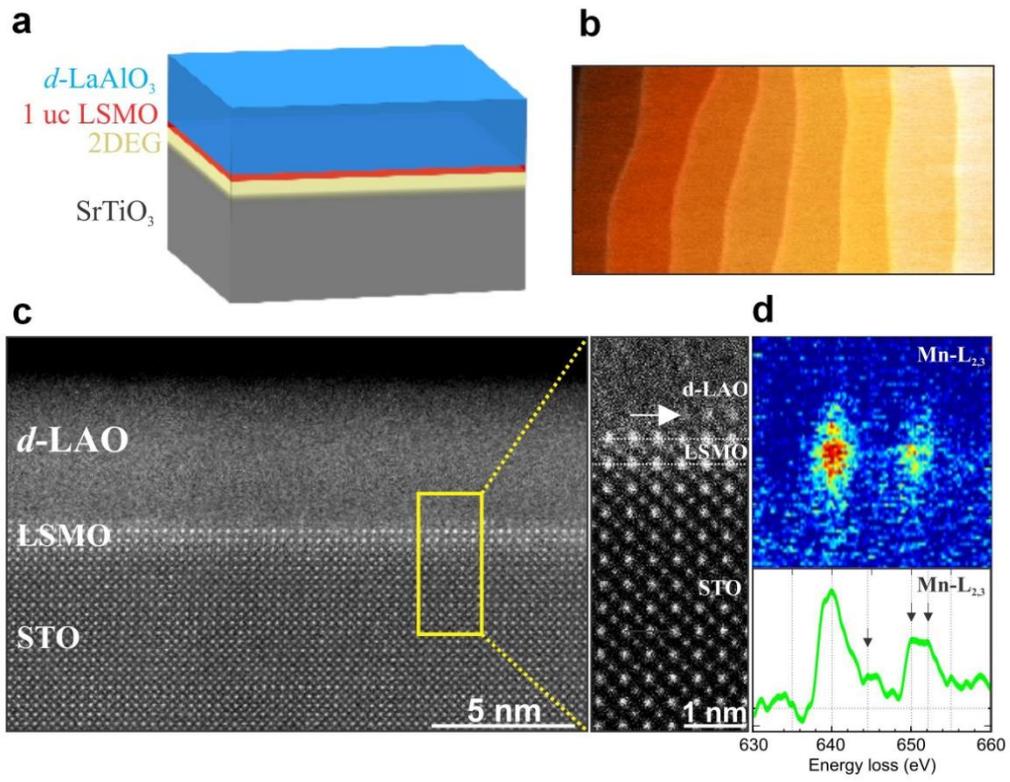

**Fig. 1.** Y. Z. Chen *et al.*

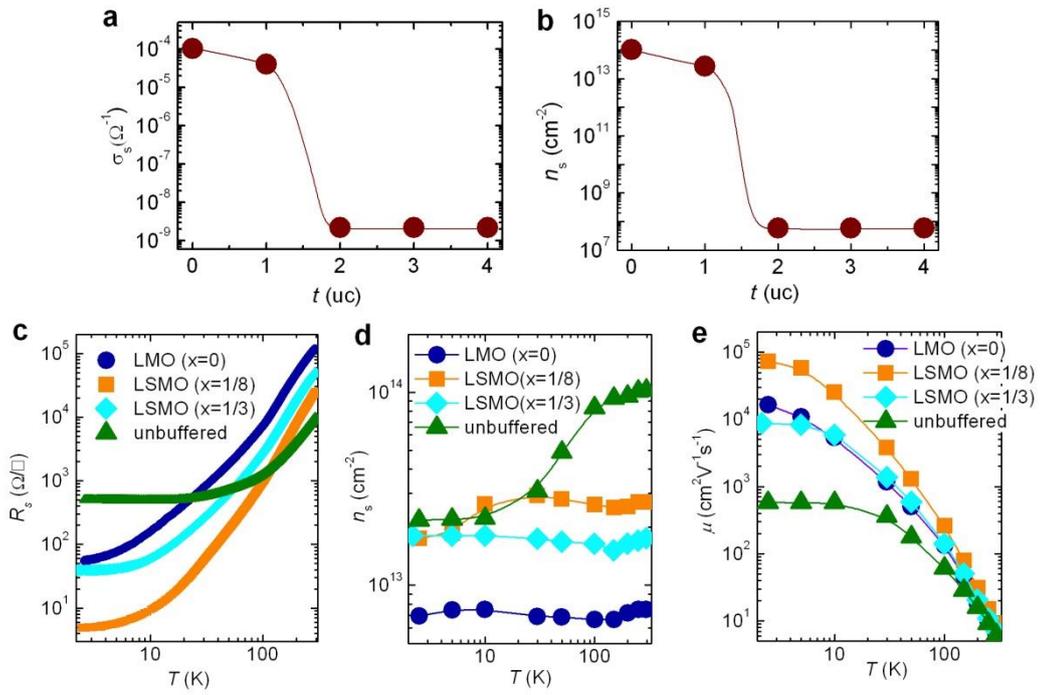

**Fig. 2.** Y. Z. Chen *et al.*

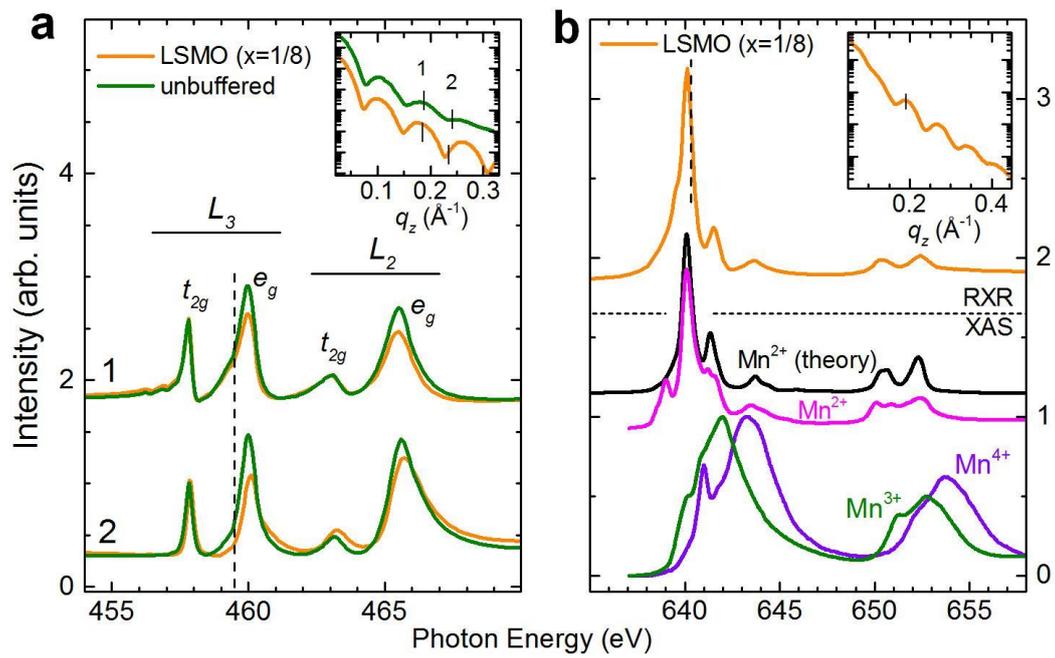

**Fig.3**. Y. Z. Chen *et al.*

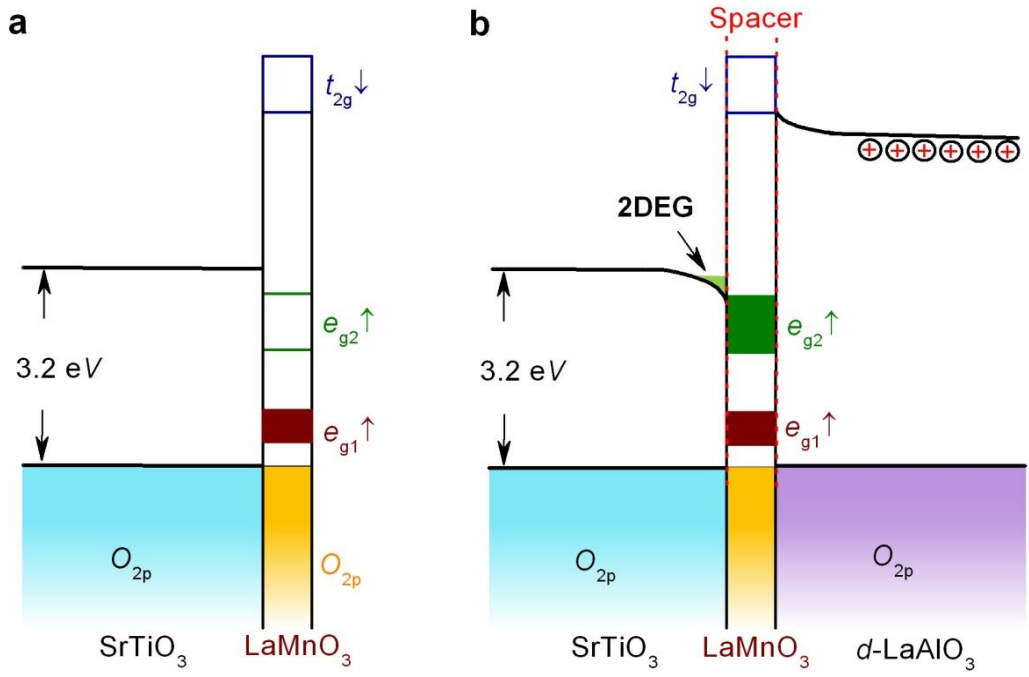

**Fig.4**. Y. Z. Chen *et al.*

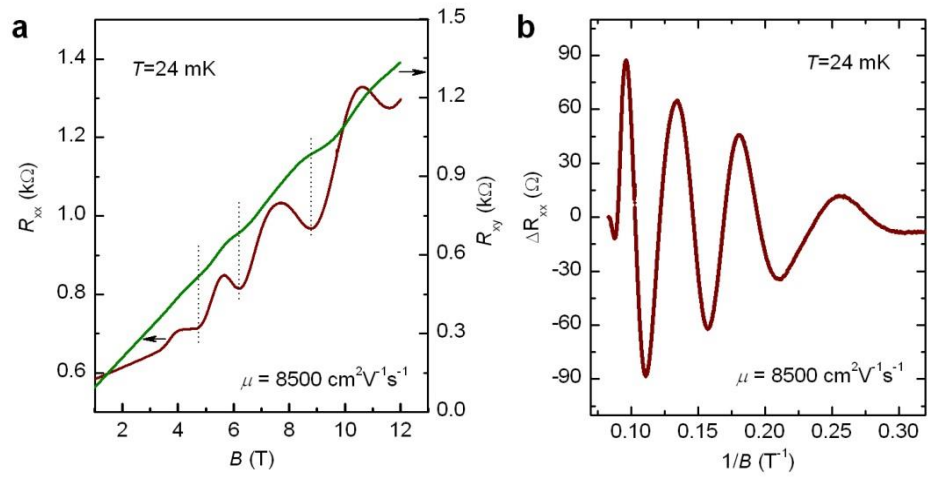

Fig. 5. Y. Z. Chen *et al.*